\documentclass[aps,amssymb,amsmath,amsfonts,twocolumn]{revtex4}
\usepackage{graphicx}
\usepackage{dsfont}
\usepackage{verbatim} 
\usepackage{mathbbol,bm,bbm}

\newcommand{\M}{\mathcal{M}}
\newcommand{\E}{\mathbb{E}}
\newcommand{\U}{\mathcal{U}}
\newcommand{\iy}{\infty}
\newcommand{\ind}{\mathbf{1}}

\DeclareMathOperator{\trace}{Tr}

\newtheorem{theorem}{Theorem}[section]

\newtheorem{lemma}[theorem]{Lemma}

\begin{document}
%\title{Ensembles of structured random states and their entanglement}
 \title{Generating random density matrices}

\medskip

\author{
Karol {\.Z}yczkowski$^{1,{2}}$, %\\[2ex]
Karol A. Penson$^{3}$,
Ion Nechita$^{4,5}$ and Beno{\^i}t Collins$^{4,6}$\\ 
\medskip
{\normalsize\itshape {$^1$%Mark Kac Complex Systems Research Centre, 
Institute of Physics,
   Jagiellonian University, ul. Reymonta 4, 30-059 Krak{\'o}w, Poland}}\\
 {\normalsize\itshape $^{{2}}$Centrum  Fizyki Teoretycznej, Polska Akademia Nauk,
   Al. Lotnik{\'o}w 32/44, 02-668 Warszawa, Poland}\\
 {\normalsize\itshape $^{{3}}$ Universit{\'e} Paris VI,
  Laboratoire de Physique de la Mati{\`e}re Condens{\'e}e (LPTMC), CNRS
  UMR 7600,  t.13, 5{\`e}me {\'e}t. BC.121, 4, pl. Jussieu, F 75252 Paris Cedex 05, France 
%Theorique des Liquides, 75252 Paris, France 
}\\
 {\normalsize\itshape $^{{4}}$Department of Mathematics and Statistics, 
University of Ottawa, ON K1N6N5, Canada}\\
 {\normalsize\itshape $^{{5}}$
Laboratoire de Physique Th{\'e}orique du CNRS, IRSAMC, Universit{\'e} de Toulouse,
 UPS, F-31062 Toulouse, France}\\
 {\normalsize\itshape $^{{6}}$CNRS, Institut Camille Jordan Universit\'e  
          Lyon 1, 43 Bd du 11 Novembre 1918, 69622 Villeurbanne, 
France}\\[2ex]}

%e-mail:

\email{karol@tatry.if.uj.edu.pl;  penson@lptl.jussieu.fr; inechita@uottawa.ca; bcollins@uottawa.ca}

\medskip

 \date{{March 18, 2011}}

\begin{abstract} 
We study various methods to generate ensembles of 
random density matrices of a fixed size $N$,
obtained by partial trace of  pure states on composite systems.
Structured ensembles of random pure states, 
invariant with respect to local unitary transformations
are introduced. To analyze statistical properties of quantum entanglement 
in bi-partite systems we analyze the distribution of 
Schmidt coefficients of random pure states.
Such a distribution is derived in the case 
of a superposition of $k$ random maximally entangled states. 
For another ensemble,  obtained by performing selective measurements
in a maximally entangled basis on a multi--partite system,
we show that this distribution is given by the Fuss-Catalan law
and find the average entanglement entropy. 
A more general class of structured ensembles proposed, 
containing also the case of Bures,
forms an extension of the standard ensemble of structureless random pure states, 
described asymptotically, as $N \to \infty$, by the Marchenko-Pastur distribution.
\end{abstract}

\maketitle

\section{Introduction}

\medskip

Random states are often used in various problems 
in quantum mechanics and the theory of quantum information.
On one hand, to describe properties of a quantum state
affected by noise present in the system, one may assume
that a given state is subjected to random interaction.
On the other hand, random quantum states emerge in a natural way
due to time evolution of arbitrary initial states of  
quantum analogues of classically chaotic systems \cite{Ha00}.
Furthermore, not knowing much about a given physical state
one can ask about their generic properties, characteristic of
a 'typical' state. 
For instance,  a key conjecture in the theory of quantum information 
concerning the additivity of minimal output entropy 
was recently shown to be false by investigating properties of random states
obtained by random operations applied to the 
maximally entangled state of a composed system \cite{Ha09}.

A standard ensemble of random pure states of size $N$
is induced by the Haar measure over the unitary group 
$U(N)$. The same construction works for quantum composite systems.
For instance, random pure states of an $N\times K$ quantum system
corresponds to the natural Fubini-Study measure,
invariant with respect to the unitary group $U(NK)$.
Thus such ensembles of random pure states are structureless,
as the probability measure is determined by the total dimension
of the Hilbert space and is does not depend on the tensor 
product structure \cite{Br96,ZS01}.

In this work we are going to analyze {\sl structured ensembles}
of random states on a composite systems, 
in which such a tensor product structure plays a crucial role.
For instance, in the case of an $N \times K$ system, 
it is natural to consider ensembles of random states
invariant with respect to local unitary transformations, 
described by the product unitary group, $U(N) \times U(K)$.
Well known constructions of random product states
and random maximally entangled states can thus serve as
simplest examples of the structured ensembles of random states.
Other examples of structured ensembles of random pure states,
which correspond to certain graphs were recently 
studied in \cite{CNZ10}.

The main aim of this paper is to introduce
physically motivated ensembles of structured random states
and to characterize the quantum entanglement of such states.
For this purpose we analyze the spectral density $P(x)$
of the reduced density matrix of size $N$, 
where $x$ denotes the rescaled eigenvalue, $x=N\lambda$.
In some cases we evaluate also the average entanglement entropy,
% $E(|\psi\rangle)$, 
defined as the Shannon entropy of the vector $\vec \lambda$  of 
the Schmidt coefficients.
%%  $\{\lambda_i\}$, $i=1,\dots, N$ of the state $|\psi\rangle$.

We treat in detail two cases of a direct importance in the
theory of quantum information.
The first ensemble is obtained by taking a coherent superposition of 
a given number of $k$ independent, random maximally entangled states
of an $N\times N$ system.
We derive explicit formulae for the asymptotic spectral density $P_k(x)$
of the reduced state and show that in the limit of large $k$, the density
converges to the {\sl Marchenko-Pastur distribution} (MP) \cite{MP67,Fo10}.
This feature is characteristic to the case of the structureless ensemble of
random pure states, which induces probability 
measures in the set of mixed quantum states \cite{Pa93,ZS01,BZ06}.

To introduce the second example of a structured ensemble
we need to consider a four partite system.
We start with an arbitrary product state
and by allowing for a generic bi-partite interaction
we create two random states on subsystems $AB$ and $CD$.
The key step is now to perform an orthogonal selective measurement
in the maximally entangled basis on subsystems
$B$ and $C$. Even though the resulting pure state on 
the remaining subsystems $A$ and $D$ does depend on the 
outcome of the measurement, its statistical properties do not,
and we show that the corresponding level density is 
given by the Fuss-Catalan distribution. 

Note that the above two constructions could be experimentally accessible, 
at least in the two-qubit case for $k=2$.
The second construction can  be easily generalized for 
a system consisting of $2s$ subsystems. An initially product state is then transformed
by a sequence of bi-partite interactions into a product of $s$ bi-partite random states.
Performing an orthogonal projection into a product of $(s-1)$ maximally entangled bases
we arrive with a random bipartite state of the structured ensemble
defined in this way. The distribution of its Schmidt coefficients is shown
to be asymptotically described by the 
{\sl  Fuss-Catalan distribution} of order $s$,
since its moments are given by the generalized Fuss-Catalan numbers
\cite{armstrong,banica-etal,mlotk}.
This distribution can be considered as a generalization 
of the MP distribution which is obtained for $s=1$.
We conclude this paper proposing a generalized, two-parameter ensemble of structured random states,
which contains, as particular cases, all the ensembles analyzed earlier in this work.

\medskip

This paper is organized as follows.
 In section II we recall necessary definitions 
% and describe ensembles product measures in the space of quantum mixed states
and describe a general scheme of generating mixed states by
taking random pure states on a composite system of a given ensemble 
and taking an average over selected subsystems.
In Section III we use $2$ independent 
random unitary operators to construct the arcsine ensemble 
obtained by superposition of two random maximally entangled states.
More general structured ensembles are obtained by superposing
$k$ random maximally entangled states.
Other ensembles of random pure states are discussed in section IV.
For completeness we review here some older results
on ensembles which lead to the Hilbert-Schmidt and the 
Bures measures in the space of density matrices.

In section V we present a scheme to generate random states
by considering a state defined on a $2s$--partite system
and performing measurement onto the product of $(s-1)$
maximally entangled states.
The distribution of the Schmidt coefficients of the resulting random pure state 
is then given by the Fuss-Catalan distribution of order $s$.
An explicit form of this distribution 
is presented in Sec. VI for any $s$.
In Sec. VII we list  various ensembles of random states,
 characterize their statistical properties 
and introduce a  generalized ensemble of structured states
(\ref{genera}, \ref{gensemble}).
% which contains previously analyzed ensembles as particular cases.
Details concerning the properties of the sum of $k$ independent
random unitary matrices are presented in Appendix A.
Brief discussion of real random pure states and
real density matrices is
provided in Appendix B, in which we state that the distribution of
singular values of a product of $s$ real random Ginibre matrices
can be asymptotically described by the Fuss--Catalan distribution.

\section{Ensembles of random states}

To construct an ensemble of random states of a given size $N$
one needs to specify a probability measure in the set of all
density operators of this size. 
 Interestingly, there is no single, distinguished probability measure
in this set, so  various ensembles of random density operators are used.

A possible way to define such an ensemble 
is to take random pure states of a given ensemble 
on bi-partite system and to average over a chosen subsystem.
The structureless ensemble of random pure states 
on $N \times N$ system distributed according to unitarily invariant measure 
leads then to the Hilbert-Schmidt measure \cite{Br96,ZS01}.
A similar construction to generate
random states according to the Bures measure  \cite{Ha98,Sl99b,BZ06}
was recently proposed in \cite{OSZ10}.
These algorithms to generate random quantum states
are already implemented in a recent {\sl Mathematica}
package devoted to Quantum Information \cite{MPG10,Mi11}.

%%%%%%%%%%%% 

Let us recall here the necessary notions and  definitions.
Consider the set of quantum states ${\cal M}_N$ 
which contains all Hermitian, positive operators $\rho=\rho^{\dagger}\ge 0$ of size
$N$ normalized by the trace condition ${\rm Tr \rho}=1$.
Any hermitian matrix can be diagonalized, $\rho= V\Lambda V^{\dagger}$.
Here $V$ is a unitary matrix consisting of eigenvectors of $\rho$,
while $\Lambda$ is the diagonal matrix containing the eigenvalues 
$\{\lambda_1, \dots, \lambda_N\}$.

In order to describe an ensemble of random density matrices, one needs to specify a 
concrete probability distribution in the set ${\cal M}_N$ of quantum states.
In this work we are going to analyze ensembles of random states, 
for which the probability measure has a product form and may be factorized \cite{Ha98,ZS01},
\begin{equation}
  {\rm d} \mu_{\rm \rho} \ =\  {\rm d} \nu
 (\lambda_1,\lambda_2,...,\lambda_N) \times {\rm d} \mu_V,
  \label{product}
\end{equation}
so the distribution of eigenvalues and eigenvectors are independent.
It is natural to assume that the eigenvectors
are distributed according to the unique, unitarily invariant,
Haar measure  $d\mu_V$  on $U(N)$. Taking this assumption as granted,
the measure in the space of density matrices will be determined
by the first factor ${\rm d} \nu$  describing the
joint distribution of eigenvalues $P(\lambda_1,\dots, \lambda_N)$.

In quantum theory mixed states arise due to an interaction of the system
investigated with an external environment. One may then 
make certain assumptions concerning the distribution of pure states 
describing the extended system. The desired mixed state $\rho$
on the principal system $A$ of size $N$ can be then obtained
by partial trace over the subsystem $B$ of an arbitrary size $K$.

Consider an arbitrary orthonormal product basis 
$|i\rangle \otimes |j\rangle \in {\cal H}_N \otimes {\cal H}_K$,
with $i=1,\dots, N$ and $j=1,\dots, K$. 
Any pure state $|\psi\rangle$ of the bi--partite system 
(not necessarily normalized)
can be expanded in this base, 
\begin{equation}
|\psi\rangle = \sum_{i=1}^N \sum_{j=1}^K X_{ij} |i\rangle \otimes |j\rangle ,
\label{randc}
\end{equation}
where $X$ is a given complex rectangular matrix of size $N \times K$.

Let us then consider an arbitrary 
complex matrix $X$. It  leads to a (weakly) positive matrix
$XX^{\dagger}$.
Thus normalizing it one obtains a legitimate quantum state,
which corresponds to the partial trace of the initial pure state
$|\psi\rangle$ over the subsystem $B$, 
\begin{equation}
\rho \ = \ \frac{ {\rm Tr}_B |\psi\rangle \langle \psi|} 
  {\langle \psi|\psi\rangle }
=
\frac{XX^{\dagger}} {{\rm Tr} XX^{\dagger}} \ .
\label{randww}
\end{equation}
For instance, taking $X$ from the {\sl Ginibre ensemble} \cite{Gi65,Me04}
 of complex square matrices we get
the Hilbert--Schmidt ensemble of quantum states, 
while a more general family of induced measures corresponds \cite{SZ04}
to the ensemble of rectangular Ginibre matrices.

The spectrum of a  density matrix $\rho$ is thus 
equivalent to the set of 
Schmidt coefficient of the initially pure state $|\psi\rangle$,
which are equal to squared singular values of a matrix $X$,
normalized in such a way that Tr$\rho=1$.
The degree of mixing of the reduced matrix $\rho$
can be characterized by its von Neumann entropy, 
$S(\rho)=-{\rm Tr} \rho \ln \rho$,
equal to the Shannon entropy of the Schmidt vector, 
$E(|\psi\rangle)=-\sum_i \lambda_i \ln \lambda_i$. This quantity
is also called {\sl entropy of entanglement}
of the pure state $|\psi\rangle$, as it is equal to zero iff the state 
 has a tensor product structure and is separable.
We are also going to use the Chebyshev entropy, 
which depends only on the largest eigenvalue
$S_{\infty}(\rho)=- \ln \lambda_{\rm max}$, 
and determines the geometric measure of entanglement
of the pure state $|\psi\rangle$.
 
Different assumptions concerning the distributions 
of pure states of the bi-partite system lead 
to different ensembles of mixed states on the system $A$. 
A natural assumption  that $|\psi\rangle$ belongs to
the standard, structureless ensemble of random pure state
distributed with respect to  the  unitarily invariant measure  
leads to the induced measures \cite{ZS01} in the space of mixed states.

In the subsequent section we are going to present simple examples
of structured ensembles of random states, in which the tensor structure plays a
key role. For instance, we discuss first ensembles
of random states of a bi-partite systems, 
which are invariant with respect to local unitary operations. 

\section{Two partite systems}

%\subsection{Maximally entangled pure states and Dirac distribution}
\subsection{Random separable and maximally entangled pure states}

For completeness we shall start the discussion with the somewhat trivial case
of generating random separable states. Consider an arbitrary product state
on a bi-partite system, $|0,0\rangle=|0\rangle_A \otimes |0\rangle_B$.
A local unitary operation, $U=U_A\otimes U_B$, cannot produce quantum 
entanglement, so the state defined by two random unitary matrices,
$|\psi_{AB}\rangle=U|0,0\rangle = U_A|0\rangle_A \otimes U_B|0\rangle_B$
is also separable. Hence a random separable
state is just a product of two random states, 
$|\psi_{AB}\rangle=|\psi_{A}\rangle \otimes |\psi_{B}\rangle$

Consider now a generalized Bell state defined on an $N\times N$ system $A,B$,
\begin{equation}
|\Psi^+\rangle = \frac{1}{\sqrt{N}} \; 
 \sum_{i=1}^N |i\rangle_A \otimes |i\rangle_B .
\label{Bell}
\end{equation}
As the entanglement entropy of this state is maximal,
$E(|\Psi^+\rangle)=\ln N$, this state is called maximally entangled.

\begin{figure}[htbp]
\centering
\includegraphics[width=0.43\textwidth]{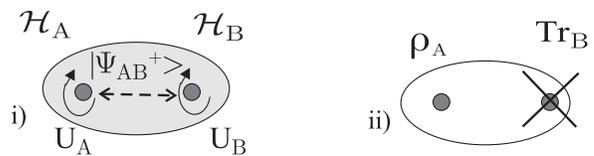}
\caption{Generating mixed states according to $\pi^{(0)}$:
 step i) a local operation $U_A\otimes U_B$ on $|\Psi_{AB}^+\rangle$
creates a random entangled state $|\phi_{\rm ent}\rangle$,
while the partial trace over an auxiliary subsystem $B$ leads in step ii) 
to the maximally mixed state on the system $A$.}
\label{fig1}
\end{figure}

Any local unitary operation, $U_A\otimes U_B$,
preserves quantum entanglement, so the locally transformed state 
$|\phi_{\rm ent}\rangle:=(U_A\otimes U_B) |\Psi^+\rangle$
remains maximally entangled. Hence taking random unitary matrices $U_A,U_B\in U(N)$
according to the Haar measure we obtain an ensemble of random entangled states, 
such that their partial trace is equal to the maximally mixed state, 
\begin{equation}
\rho={\rm Tr}_B |\phi_{\rm ent}\rangle \langle \phi_{\rm ent}| 
  =  \frac{1}{N} {\mathbb 1}_N =: \rho_* .
\label{belltrac}
\end{equation}
This Dirac distribution will be denoted by $\pi^{(0)}(\rho)=\delta(\rho-\rho_*)$
and a scheme of generating it by averaging over 
an auxiliary subsystem is shown in Fig.~\ref{fig1}.

\subsection{Arcsine ensemble}

\begin{figure}[htbp]
\centering
\includegraphics[width=0.43\textwidth]{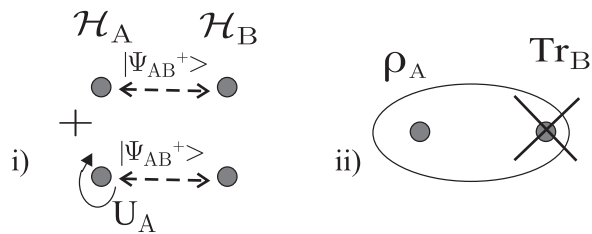}
\caption{To generate states from the arcsine ensemble described by (\ref{cosin})
one has to i) construct a superposition of a maximally entangled state $|\Psi_{AB}^+\rangle$
with another maximally entangled state 
$(U_A \otimes {\mathbbm 1}) |\Psi_{AB}^+\rangle$, 
and ii) perform partial trace over an auxiliary subsystem $B$.}
\label{fig2}
\end{figure}

Consider now a more general case of a coherent superposition
of two maximally entangled states. To be precise
we fix a given maximally entangled state $|\psi_1\rangle$,
and use such a local basis that it is represented by (\ref{Bell}).
Taking a local random unitary matrix $U_A\in U(N)$
we generate another maximally entangled state 
$|\psi_2\rangle=U_A\otimes {\mathbbm 1}|\psi_1\rangle$.
As shown in Fig.~\ref{fig2}
we construct their symmetric superposition, 
\begin{equation}
|\phi\rangle = 
  ( |\psi_1\rangle +|\psi_2\rangle)=
  [{\mathbbm 1}_{N^2} +(U\otimes {\mathbbm 1}_N)]
 |\Psi^+\rangle \ .
\label{comb1}
\end{equation}
Note that this ensemble is invariant
 with respect to local unitary operations, $U(N)\times U(N)$.
Let us specify a subsystem $B$
and average over it to obtain the reduced state 
\begin{equation}
\rho \ = \ 
\frac{ {\rm Tr}_B |\phi\rangle \langle \phi |}
{\langle \phi|\phi\rangle} 
= \frac{ 2{\mathbbm 1}+U+U^\dagger}{2N +{\rm Tr}(U+U^{\dagger})} .
\label{cosrand}
\end{equation}

If the matrix $U$ is generated according to the Haar measure on $U(N)$
its eigenphases 
are distributed according to the uniform distribution \cite{Me04},
$P(\alpha)=1/2\pi$ for $\alpha\in [0,2\pi)$.
Thus the term ${\rm Tr}(U+U^{\dagger})$
present in the normalization constant becomes negligible for large $N$.
Therefore  eigenvalues of a random density matrix $\rho$
by  (\ref{cosrand}) have the form $\lambda_i=(1+\cos\alpha_i)/N$,
for $i=1,\dots, N$.
Making use of the rescaled variable 
$x=\lambda N$ we arrive at a conclusion that the 
spectral level density of random density matrices (\ref{cosrand})
is asymptotically described by the {\sl arcsine} distribution \cite{HP00} 
\begin{equation}
P_{\rm arc}(x) = \frac{1}{\pi \sqrt{x(2-x)}} 
\label{cosin}
\end{equation}
defined on the compact support $[0,2]$.
This is a particular case of the beta distribution 
(with parameters $\alpha=\beta=1/2$) and the Dirichlet distribution.
It describes the Jeffreys prior of a Bernoulli trial \cite{Hy06} 
and is related to the statistical 
distance between classical probability distributions \cite{BZ06}.
The name  `arcsine' is due to the fact that the 
corresponding cumulative distribution
is proportional to $\sin^{-1}\sqrt{x/2}$.
Thus the ensemble of random density matrices constructed according to the
procedure shown in  Fig.~\ref{fig2} will be called {\sl arcsine ensemble}.
The average entropy for the arcsine distribution reads 
$\int_0^2  -x \ln x P_{\rm arc}(x) =\ln 2-1\approx -0.307$,
so the average entropy of entanglement of a random pure state
on the $N\times N$ system generated 
form this ensemble reads $\langle E\rangle_{\psi}\approx \ln N -1+\ln2.$

\smallskip

 It is clear that one may define a family of interpolating ensembles
by taking a convex combination of the mixed states defined by 
(\ref{belltrac}) and  (\ref{cosrand}).
In other words, one can take a family of random matrices 
parametrized by a real number $a\in [0,1]$
and write $W_a=a{\mathbbm 1}+(1-a)U$.
Plugging this expression in place of $X$ into  (\ref{randww})
we construct a family of ensembles of density matrices
which gives the Dirac mass for $a=0$ and $a=1$,
while the arcsine ensemble is obtained for $a=1/2$.

\subsection{Superposition of $k$ maximally entangled states}

The arcsine ensemble introduced above can be obtained by 
superimposing $k=2$ random maximally entangled states.
It is straightforward to generalize this construction for
an arbitrary number of $k$ maximally entangled states.
Each of them can be written by an action of a local unitary
on the fixed maximally entangled state  $|\Psi^+\rangle $.
More precisely we set 
$|\psi_i\rangle=(U_i \otimes {\mathbbm 1}) |\Psi^+\rangle $
for $i=1,\dots, k$ and construct their equi-probable superposition
$|\phi\rangle = \sum_{i=1}^k |\psi_i\rangle=
 \bigl[(\sum_{i=1}^k U_i) \otimes {\mathbbm 1}_N \big] |\Psi^+\rangle $.

As before this ensemble is invariant with respect to local transformations.
The random mixed state is obtained by taking 
the partial trace over the subsystem $B$  and normalizing the outcome
$\rho  =  
 {\rm Tr}_B |\phi\rangle \langle \phi | /
 {\langle \phi|\phi\rangle}$ 
as in eq. (\ref{cosrand}).
This procedure leads now to the following 
expression for a mixed state, generated by $k$ 
independent random unitaries, 
\begin{equation}
\rho \ = \ 
 \frac{ (U_1 + \dots +U_k)(U_1^{\dagger} + \dots +U_k^{\dagger})}
{{\rm Tr} (U_1 + \dots +U_k)(U_1^{\dagger} + \dots +U_k^{\dagger})} .
\label{cosrandk}
\end{equation}

As shown in Appendix A the spectral density of
random states defined above converges, for large system size $N$
to the following distribution $\nu_k(x)$
defined for $x\in [0, 4\frac{k-1}{k}]$,
\begin{equation}
\label{eq:density-k-rescaled}
	\nu_k(x) = \frac{1}{2\pi} \frac{\sqrt{4k(k-1)x - k^2x^2}}{kx - x^2}
\end{equation}

\begin{figure}[htbp]
\centering
\includegraphics[width=0.40\textwidth]{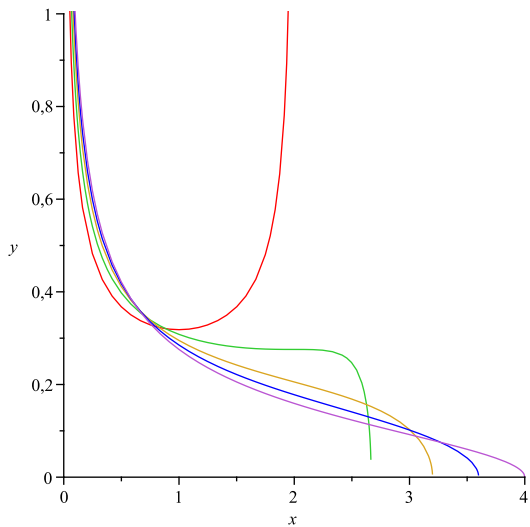}
\caption{Family of spectral distributions $\nu_k(x)$
plotted for $k=2,3,5,10$ and $k\to \infty$.
The support of the measures is an increasing 
function of the parameter $k$.
They describe level density of mixed states
generated by superposition of $k$ 
random maximally entangled states
and performing the partial trace.
 The case $k=2$ corresponds to the
arcsine ensemble, while in the case $k\to \infty$
the sum of $k$ random unitaries has properties of a random Ginibre matrix,
so the spectral distribution converges to the Marchenko-Pastur law.}
\label{fig3b}
\end{figure}

The shape of these distributions is presented 
for some values of $k$ in Fig.  \ref{fig3b}.
It is easy to see that for a large number $k$ the above measures converge weakly to the limit
$\nu_\infty = \pi^{(1)} = \frac{\sqrt{4x-x^2}}{2\pi x}$
which is the Marchenko-Pastur distribution $\pi^{(1)}$.
In other words, the superposition of a large number of
random maximally entangled states destroys the structure of the ensemble, 
as the resulting state becomes typical of  the structureless ensemble.

The  measure  (\ref{eq:density-k-rescaled})
 can be connected to the free Meixner measures
 of \cite{bozejko-bryc}:
 $\nu_k$ is the free Meixner measure of 
parameters $a=0$ and $b = -1/k$.
The measure $\nu_k(x)$ can be characterized by its second moment, 
which gives the asymptotic average purity of random mixed states 
obtained by a superposition of $k$ maximally entangled states, 
$\langle {\rm Tr} \rho^2\rangle_{\nu_k}\approx (2-1/k)\frac{1}{N}.$

\subsection{Induced measures}

\begin{figure}[htbp]
\centering
\includegraphics[width=0.43\textwidth]{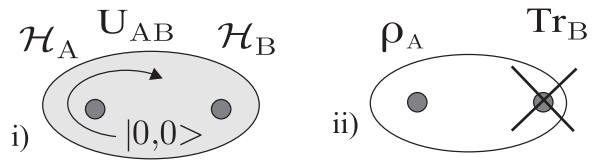}
\caption{Taking i) a random pure state $U_{AB}|0,0\rangle$
on a bi-partite system $AB$
and performing ii) the partial trace over subsystem $B$ 
leads random state distributed according to an induced measure,
with spectral density described by the Marchenko-Pastur distribution.}
\label{fig3}
\end{figure}

Consider a quantum system composed of two subsystems. 
Let $N$ denote the dimension of the principal system, which 
interacts with a $K$ dimensional environment.
Assume first that the state $|\psi\rangle\in {\cal H}_N \otimes {\cal H}_K$
is taken from the structureless ensemble, 
so it is distributed uniformly according to the Fubini--Study measure
on the complex projective space. In other words, the random pure state
can be represented by $|\psi\rangle=U|0,0\rangle$, where
$U$ is a global random unitary matrix distributed according to the Haar measure 
-- see Fig. \ref{fig3}. The initial state is arbitrary,
so for concreteness we may choose it as a given product state $|0,0\rangle$ 
 in ${\cal H}_N \otimes {\cal H}_K$.
The mixed state $\rho$ obtained by the partial trace
over the $K$ dimensional environment
\begin{equation}
\rho \ = \  {\rm Tr}_K |\psi\rangle \langle \psi | \; ,
\quad {\rm with \quad}
|\psi\rangle \in {\cal H}_N \otimes {\cal H}_K ,
\label{hsrand2}
\end{equation}
is distributed according to the measure $\mu_{N,K}$
in the space of density matrices of size $N$,
induced by the Haar measure on the unitary group $U(NK)$.
This procedure yields the following probability distribution
\begin{equation}
    d\mu (\rho) \ \propto \
\Theta (\rho) \ \delta({\rm Tr} \rho -1) \det \rho^{K-N}.
\label{dmr}
\end{equation}
The $\Theta$ step--function and the Dirac $\delta$
reflect key properties of density matrices:
positivity, $\rho \ge 0$ and normalization,
${\rm Tr} \rho =1$.

If $|\psi\rangle$ is a random pure state distributed uniformly according to the
unitarily invariant Fubini--Study measure in the space 
of pure states  all elements of $G$ are independent complex Gaussian variables
with the same variance,
so this matrix belongs to the Ginibre ensemble \cite{Gi65}.
Hence the reduced state (\ref{hsrand2}) obtained by partial trace
has the form (\ref{randww}) with $X=G$.  The eigenvalues of a random matrix 
$\rho$ generated with respect to the measure $\mu_{N,K}$
are thus equal to the squared singular eigenvalues of a normalized
rectangular $N \times K$ complex random matrix $G$ from the Ginibre ensemble.

Any density matrix is Hermitian and can be diagonalized by a unitary rotation.
Integrating out the eigenvectors of a random state $\rho$ defined by  (\ref{dmr})
one reduces $d\mu$ to the measure on the simplex of eigenvalues $\{ \lambda_1, \ldots, \lambda_N\}$ 
of the density operator \cite{Lu78},
\begin{equation}
\label{dmuM}
P_{N,K}(\lambda_1,...\lambda_N)  = C_{N,K}
  \prod_i  \lambda_i^{K-N}
\prod_{i<j}^{1...N} (\lambda_i -\lambda_j )^2  \ .
\end{equation}
In this expression all eigenvalues are assumed to be non-negative, $\lambda_i\ge 0$
and they sum to unity,  $\sum_i\lambda_i =1$.
 The normalization constant can be expressed \cite{ZS01} by the Gamma function  
\begin{equation}
C_{N,K} = \frac{\Gamma(KN)} {\prod_{j=0}^{N-1} \Gamma(K-j) \Gamma(N-j+1) } .
\label{cnm}
\end{equation}

Induced measures are stable with respect to the partial trace,
 what can be formulated as follows.

{\bf Proposition 1}. Consider a random state 
$\rho$ generated according to the induced measures $\mu_{M,K}$.
Assume that the dimension is composed,  $M=N \times L$, and define the partial trace
over the $L$ dimensional subsystem, $\sigma={\rm Tr}_L (\rho)$.
Then the reduced matrix $\sigma$ is generated according to the measure
$\mu_{N,KL}$.

{\bf Proof.} The state $\rho$ can be considered as a random  pure state
 $|\psi\rangle \in {\cal H}_M \otimes {\cal H}_K$ averaged over 
$K$ dimensional environment. Averaging the projector $|\psi\rangle \langle \psi|$
over a composed subsystem of size  $KL$ we arrive at state $\sigma$ of size $N$,
which shows \cite{CSZ07} that it is distributed according to 
the induced measure $\mu_{N,KL}$.

\section{Measures defined by a metric}

Let $d$ denote any distance in the space of the normalized
density operators of a fixed size $N$. With respect to this distance 
one can define unit spheres and unit balls. Attributing the same weight
to any ball of radius equal to unity one defines the measure 
corresponding to a given distance.

Consider, for instance, the Hilbert-Schmidt distance between any two mixed states $D_{HS}(\rho,\sigma)=[{\rm Tr}(\rho-\sigma)^2]^{1/2}$.
This distance induces in the space of density operators an Euclidean geometry:
the space of one--qubit mixed states has a form of the three-ball bounded by the Bloch sphere.

\subsection{Hilbert--Schmidt ensemble}

The Hilbert--Schmidt measure defined in this way 
belongs to the class of induced measures \cite{ZS01}
and can be obtained by a reduction
of random pure states defined on a bi--partite quantum system.
Looking at expression (\ref{dmuM})
we see that in the special case $K=N$  the term with the determinant
in is equal to unity, so measure reduces to the Hilbert-Schmidt measure \cite{ZS01}.
This observation leads to a simple algorithm to generate a
Hilbert-Schmidt random matrix \cite{ZS01}:
i) Take a square complex random matrix $A$ of size $N$ pertaining to the
Ginibre ensemble \cite{Gi65,Me04} (with real and imaginary parts of
each element being independent normal random variables);
ii) Write down the random matrix
\begin{equation}
\rho_{\rm HS} \ = \  \frac{\ \ GG^{\dagger}} {{\rm Tr}GG^\dagger} ,
\label{hsrand}
\end{equation}
which is by construction Hermitian, positive definite
and normalized, so it forms a legitimate density matrix.

To characterize spectral properties of random density matrices
one considers the joint distribution of eigenvalues (\ref{dmuM}),
integrates out $N-1$ variables $\lambda_1\dots \lambda_N$
to obtain the level density denoted by $P(\lambda)$.
This problem was discussed by Page \cite{Pa93},
who found the asymptotic distribution
for the rescaled variable $x=N\lambda$.
The result depends on the ratio of both dimensions $c=K/N$ 
and is given by the Marchenko--Pastur distribution \cite{MP67},
\begin{equation}
\pi^{(1)}_c(x) =
\max(1-c,0) \delta(0)+
 \frac{\sqrt{4c-(x-c-1)^2}}{2 \pi x} .
\label{densMPc}
\end{equation}
This expression is valid for $ x\in [x_-,x_+]$,
where $x_{\pm}=1+c\pm 2\sqrt{c}$.
In the case $c=1$ corresponding to the Hilbert--Schmidt ensemble 
this distribution reads  
\begin{equation}
P_{\rm HS}(x)=  \frac{1}{2\pi}
 \sqrt{\frac{4}{x} -1}
{\quad \rm for \quad}
x\in [0,4]
\label{densitHS}
\end{equation}
It diverges as $x^{-1/2}$ for $x\to 0$ and 
 in the rescaled variable, $y=\sqrt{x}$
it becomes a quarter--circle law,
$P(y)=\frac{1}{\pi}\sqrt{4-y^2}$.

The average von Neumann entropy of random states
distributed with respect to the HS measure reads \cite{Pa93,SZ04}
\begin{equation}
  \langle S(\rho) \rangle_{\rm HS} =
   \ln N - \frac{1}{2} +
   O\left( \frac{\ln N}{N} \right)  .
\label{entrHS}
 \end{equation}
As the rescaled eigenvalue is $x=N\lambda$, 
this result is consistent with the fact that 
the average entropy of the asymptotic distribution (\ref{densitHS})
reads 
$\int_0^4 -x \ln x P_{\rm HS}(x) dx = -1/2$.
Note that the mean entropy of a random mixed state is close to the maximal
entropy in the $N$ dimensional system, which equals $\ln N$
for the maximally mixed state $\rho_*={\mathbbm 1}/N$.
The asymptotic average purity $\langle {\rm Tr} \rho^2\rangle_{HS}\approx 2/N$
which is consistent with the second moment of the MP distribution, 
$\int_0^4 x^2 P_{\rm HS}(x)=2$.

\subsection{Bures ensemble}

Another distinguished measure in the space $\Omega$ of quantum mixed states,
is induced by the Bures distance \cite{Bu69,Uh92},
\begin{equation}
D_B(\rho,\sigma)=
\sqrt{ 2- 2{\rm Tr}(\sqrt{\rho}\sigma \sqrt{\rho})^{1/2}} .
\label{bures1}
\end{equation}
The Bures distance plays an important role in the investigation of the set 
of quantum states \cite{BZ06}.
The Bures metric, related to quantum distinguishability \cite{Hy06},
is known to be the minimal monotone metric \cite{PS96}
and applied to any two diagonal matrices it gives their statistical distance.
These special features support the claim
that without any prior knowledge on a certain state acting on ${\cal H}_N$,
the optimal way to mimic it is to generate a random density operator 
with respect to the Bures measure.

The Bures measure is characterized by the following joint 
probability of eigenvalues  \cite{Ha98}
\begin{equation}
\label{dmuB}
 P_B(\lambda_1,...\lambda_N) =
  C_N^B 
\prod_i \lambda_i^{-1/2} 
\prod_{i<j}^{1...N}
\frac{(\lambda_i -\lambda_j )^2}{ \lambda_i +\lambda_j} ,
\end{equation}
where all eigenvalues are non-negative, $\lambda_i\ge 0$
and they sum to unity,  $\sum_i\lambda_i =1$.
The normalization constant for this measure
\begin{equation}
C_N^B =
2^{N^2-N}\ \frac{\Gamma(N^2/2)}
  {\pi^{N/2}\  \prod_{j=1}^{N} \Gamma(j+1) }
\label{burcn}
\end{equation}
was obtained in \cite{Ha98,Sl99b} for small  $N$ and in \cite{SZ03}
in the general case.

To generate random states with respect to the Bures ensemble
one can proceed according to the following algorithm \cite{OSZ10}:

i) Take a complex random matrix $G$ of size $N$ pertaining to the
 Ginibre ensemble and a random unitary matrix $U$
  distributed according to the Haar measure on $U(N)$ \cite{PZK98}.

ii) Write down the random matrix
\begin{equation}
\rho_{\rm B} \ = \  \frac{\ \ ({\mathbbm 1}+U)GG^{\dagger} ({\mathbbm 1}+U^{\dagger})}
{{\rm Tr}[({\mathbbm 1}+U)GG^{\dagger} ({\mathbbm 1}+U^{\dagger})]} \,
\label{burrand}
\end{equation}
which is distributed according to the Bures measure.
In the analogy to the Hilbert-Schmidt ensemble we can write
this state as reduction of a pure state on the composed system, 
\begin{equation}
\rho_{\rm B} \ = \ \frac{{\rm Tr}_N |\phi\rangle \langle \phi |}
{\langle \phi|\phi\rangle }
\quad {\rm where \quad}
|\phi\rangle := [({\mathbbm 1}+V_A)\otimes {\mathbbm 1}]|\psi_1\rangle \; .
\label{burrand2}
\end{equation}
Here $|\psi_1\rangle=U_{AB}|0,0\rangle$ 
is a random state on the bipartite system used in  Eq. (\ref{hsrand2})
and $V_A\in U(N)$.
\begin{figure}[htbp]
\centering
\includegraphics[width=0.43\textwidth]{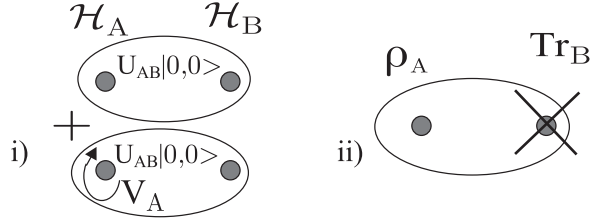}
\caption{To generate states according to the Bures distribution 
one i) constructs a superposition of a random bi-partite state
 $|\psi_1\rangle=U_{AB}|0,0\rangle$
with a locally transformed random state 
$|\psi_2\rangle=(V_A \otimes {\mathbbm 1})|\psi_1\rangle$, 
and ii) perform partial trace over an auxiliary subsystem $B$.}
\label{fig4}
\end{figure}

The asymptotic probability distribution for the rescaled
eigenvalue $x=N\lambda$ of a random density matrix
generated according to the Bures ensemble reads
 \cite{SZ04}
\begin{equation}
P_{\rm B}(x)= 
C
\left[ 
\left(\frac{a}{x} +\sqrt{\left(\frac{a}{x}\right)^2 \! \! -1} \right)^{2/3}
\!\!\!\!\!   -  \!
\left(\frac{a}{x} -\sqrt{\left( \frac{a}{x}\right)^2 \!  \! -1} \right)^{2/3}
\right]
\label{densitB}
\end{equation}
where $C=1/4\pi \sqrt{3}$ and $a=3\sqrt{3}$.
This distribution is defined on a support larger than the standard MP distribution, 
$x \in [0, a]$ and it diverges for $x\to 0$ as $x^{-2/3}$.

The average entropy of a random state form the Bures ensemble reads \cite{SZ04}
 \begin{equation}
  \langle S(\rho) \rangle_{\rm B} =
   \ln N - \ln 2 +
   O\left( \frac{\ln N}{N} \right)  .
 \end{equation}
This value is smaller than the average entropy (\ref{entrHS}),
which shows that the Bures states are typically less 
mixed that the states from the Hilbert--Schmidt ensemble.
A similar conclusion follows from the comparison of 
average purity for the Bures ensemble,
$\langle {\rm Tr} \rho^2\rangle_{B}\approx 5/2N$ \cite{SZ04},
with the average purity for the HS measure.

By considering random matrices of the form
$W= (a {\mathbbm 1}+ (1-a)U)G$
one may construct a continuous family of measures 
interpolating between the Bures and the Hilbert--Schmidt 
ensembles \cite{OSZ10}
and labeled by a real parameter $a \in [0,1/2]$.
A more general class of interpolating ensembles is
proposed in Sec. \ref{sec_concl}.

\section{Projection onto maximally entangled states}

\subsection {Four-partite systems and measurements in 
         a maximally entangled basis}

Consider a system consisting of four subsystems, labeled as $A,B,C$ and $D$.
For simplicity assume here that the dimensions of all subsystems are equal, 
$N_1=N_2=N_3=N_4=N$. Consider an arbitrary four-partite product state, 
say $|\psi_0\rangle=|0\rangle_A \otimes |0\rangle_B 
\otimes|0\rangle_C \otimes|0\rangle_D =:|0,0,0,0\rangle$. 
Taking two independent random unitary matrices
$U_{AB}$ and $U_{CD}$ of size $N^2$, which act on the first
and the second pair of subsystems, respectively, we define a random state
$|\psi\rangle=U_{AB}\otimes U_{CD} |\psi_0\rangle$.
By construction, it is a product state with respect to the 
partition  into two parties: $(A,B)$ and $(C,D)$. 
In the analogy to (\ref{randc}) it can be expanded in the product basis,
\begin{equation}
|\psi\rangle = \sum_{i,j=1}^N \sum_{k,l=1}^N G_{ij} E_{kl} \;
  |i\rangle_A \otimes |j\rangle_B \otimes |k\rangle_C \otimes |l\rangle_D
\label{randgg}
\end{equation}

Consider now a maximally entangled state on the second and the third subsystem,
\begin{equation}
|\Psi_{BC}^+\rangle = \frac{1}{\sqrt{N}} \; 
 \sum_{\mu=1}^N |\mu\rangle_B \otimes |\mu\rangle_C , 
\label{maxent}
\end{equation}
and the corresponding projector $P_{BC}:=|\Psi_{BC}^+\rangle \langle \Psi_{BC}^+|$.
One can extend it into a four-partite operator and define
\begin{equation}
P:= {\mathbbm 1}_A \otimes \Bigl(
\frac{1}{N}\sum_{\mu,\nu} |\mu,\mu\rangle_{BC} \langle \nu,\nu| \Bigr)
 \otimes {\mathbbm 1}_D
\label{proj}
\end{equation}
Let us assume that the random pure state $|\psi\rangle$ 
defined in (\ref{randgg}) is 
subjected to a projective measurement performed onto the second and third subsystem, 
and that the result is post-selected to be associated to the projector $P$.
This leads to a non-normalized pure state $|\phi\rangle$ 
describing the remaining two subsystems,
\begin{equation}
|\phi\rangle = P|\psi\rangle = \frac{1}{N}
 \sum_{i=1}^N \sum_{l=1}^N \sum_{k=1}^N G_{ik} E_{kl} \; 
    |i\rangle_A  \otimes |l\rangle_D .
\label{phia}
\end{equation}
Note that the projection on the entangled state $|\Psi_{BC}^+\rangle$
introduces a coupling between the subsystems $B$ and $C$.
%%%%%%%%%%%%
A similar idea was used in analysis of entanglement swapping \cite{ZZHE93}.
and in studies of 'matrix product states' \cite{FNW92}
and 'projected entangled pair states' \cite{VC04,SPGC10}.
Constructing the matrix product states an $N\times N$ entangled state
is projected down into a subspace of an arbitrary dimension $d$,
while in our approach a projection onto the maximally entangled state of $BC$ takes place, 
(which formally corresponds to $d=1$),
and only two edge systems labeled by $A$ and $D$ survive the projection.

Normalizing the resulting state $|\phi\rangle$
and performing the partial trace over the fourth subsystem $D$
we arrive at a compact expression for the resulting mixed state
on the first subsystem $A$, 
\begin{equation}
\rho \ = \  \frac{  {\rm Tr}_D |\phi\rangle \langle \phi |}
  {\langle \phi|\phi\rangle}=
 \frac{GEE^{\dagger}G^{\dagger}} {{\rm Tr}\; GEE^{\dagger}G^{\dagger}} .
\label{prodrand}
\end{equation}
For any matrices $G$ and $E$ this expression provides a valid quantum state, 
normalized and positive.
If initial pure states are random,
then the matrices $G$ and $E$ belong to 
the Ginibre ensemble and the spectrum of a random state $\rho$ 
consists of squared singular values of the product $GE$ 
of two random Ginibre matrices.
Random states with the same statistical properties
were recently found in ensembles associated with certain graphs \cite{CNZ10}.

\begin{figure}[htbp]
\centering
\includegraphics[width=0.43\textwidth]{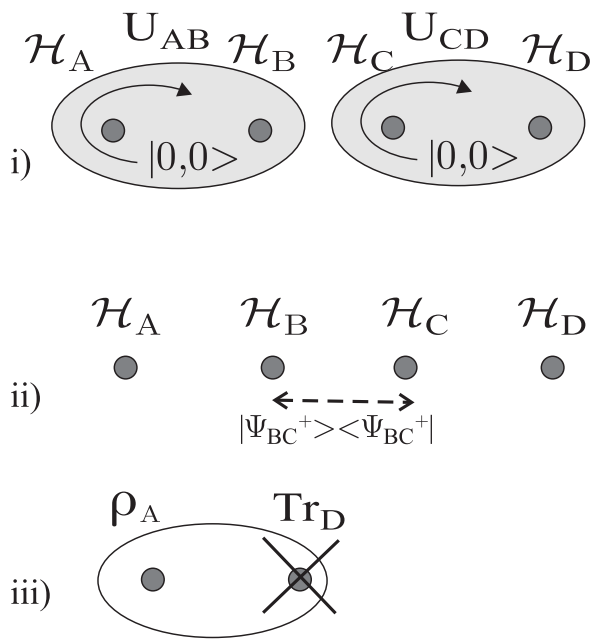}
\caption{To generate a random mixed state with spectral density
$\pi^{(2)}$ one has to i) take a product random pure state $|\psi_{AB}\rangle\otimes |\psi_{CD}\rangle$
on a four particle system $A,B,C,D$,
ii) measure subsystems  $B,C$ by a projection onto the maximally 
   entangled state $|\Psi_{BC}^+\rangle$, 
iii) average over the subsystem $D$.}
\label{fig5}
\end{figure}

Observe that the statistical properties of the ensemble defined will change
if the projection is performed with respect to an arbitrary maximally
entangled state, $|\Psi_{BC}^+\rangle'=(U_B\otimes U_C) |\Psi_{BC}^+\rangle$,
as the local unitaries $U_B$ and $U_C$ can be absorbed in the definition
of random Ginibre matrices, $G$ and $E$, respectively.

Furthermore, one may chose $N^2$ unitary matrices $U_i\in U(N)$,
which are orthogonal
in sense of the Hilbert--Schmidt scalar product, ${\rm Tr}\; U_iU_j^{\dagger}=N \delta_{ij}$.
Then the set of $N^2$ maximally entangled states, 
$|\Psi_{i}^+\rangle=(U_i\otimes {\mathbbm 1}) |\Psi_{BC}^+\rangle$,
forms a maximally entangled basis, which are known to exist in any dimension \cite{We01}.
Thus one may consider another set up in which 
a selective measurement on subsystems $B$ and $C$ is performed 
in the maximally entangled basis.
The outcome state on subsystems $AD$  depends
on the result of the quantum measurement of subsystems $BC$.
However, these results are equivalent up to a unitary transformation, which
again can be absorbed into the definition of the Ginibre ensemble.
Thus the statistical properties of the random state (\ref{prodrand})
on subsystem $A$ obtained in consequence of the measurement in the maximally entangled basis
in $BC$ followed by the partial trace over subsystem $D$
do not dependent on the outcome of the measurement.

Furthermore, the same construction holds in a more general setup
in which some dimensions of four subsystems are different.
To use the maximally entangled state $|\Psi_{BC}^+\rangle$
we set $N_B=N_C$, but the dimensions $N=N_A$ and $N_D$
can be different. This leads to two rectangular random Ginibre matrices, 
$G$ of size $N \times N_B$ and $E$ of size $N_B \times N_D$.
As in the previous case formula (\ref{prodrand}) provides a density matrix
$\rho$ of size $N$, but now the model is a function of two parameters:
dimensions $N_B$ and $N_D$. It is sometimes convenient to use two ratios, 
$c_1=N_B/N$ and $c_2=N_D/N$, so the standard version of the model
corresponds to putting $c_1=c_2=1$.

\subsection {Multi--partite systems}
Another possibility to generalize the model is to consider a larger system
consisting of an even number $2s$ of subsystems.
For simplicity assume first that their dimensions are set to $N$.
In analogy to (\ref{randgg}) we use $s$ independent unitaries 
of size $N^2$ to generate a random pure state $|\psi\rangle$.

To work with an arbitrary even number of subsystems
it is convenient to modify the notation 
and label the subsystems by integers $1,2,\dots, 2s$.
Consider an arbitrary product state of a 
$2s$ -- particle system, 
$|\psi_0\rangle=|0\rangle_1 \otimes \dots \otimes|0\rangle_{2s} =:|0,\dots,0\rangle$. 
Taking $s$ independent Haar random unitary matrices
$U_{1,2}, U_{3,4}, \dots U_{2s-1,2s}$ of size $N^2$, 
we define a random state
$|\psi\rangle=U_{1,2}\otimes \dots U_{2s-1,2s} |\psi_0\rangle$.
Expanding this state in the product basis one obtains
\begin{eqnarray}
|\psi\rangle &=& \bigl( U_{1,2}\otimes \cdots U_{2s-1,2s}\bigr)  |0,\dots ,0\rangle  \\
   &=&
 \sum_{i_1,\dots i_{2s}}
        (G_1)_{i_1,i_2} \cdots (G_s)_{i_{2s-1},i_{2s}} |i_1, \dots , i_{2s}\rangle
\nonumber
\label{randgs}
\end{eqnarray}
Performing a projection onto a product of 
$s-1$ maximally entangled states, 
\begin{equation}
P_s:= {\mathbbm 1}_1 \otimes 
         |\Psi_{2,3}^+\rangle \langle \Psi_{2,3}^+| \otimes \cdots \otimes
       |\Psi_{2s-2,2s-1}^+\rangle \langle \Psi_{2s-2,2s-1}^+|
\otimes {\mathbbm 1}_{2s}
\label{projs}
\end{equation}
we obtain a pure state $|\phi\rangle$ 
describing the remaining two subsystems,
\begin{equation}
|\phi\rangle = P|\psi\rangle = N^{1-s}
 \sum_{i,j}  
\Bigl( G_1 G_2 \cdots G_s\Bigr)_{ij}  
    |i\rangle_1  \otimes |j\rangle_{2s} .
\label{phias}
\end{equation}

Normalizing this state and performing the partial trace over the last subsystem 
we obtain an explicit expression for the resulting mixed state
on the first subsystem 
\begin{equation}
\rho \ = \  \frac{  {\rm Tr}_{2s} |\phi\rangle \langle \phi |}
  {\langle \phi|\phi\rangle}=
 \frac{G_1G_2 \cdots G_s (G_1G_2 \cdots G_s)^{\dagger}}
 {{\rm Tr}\; [ G_1G_2 \cdots G_s (G_1G_2 \cdots G_s)^{\dagger}]}
\label{prodrands}
\end{equation}
Alternatively, one may assume that this state is obtained as a results of
an orthogonal measurement into the product of $s-1$ maximally entangled
bases. The first entangled basis correlates subsystem $2$ with subsystem $3$, 
the next one couples subsystem $4$ with $5$, while due to the last one 
the  subsystem $2s-2$ is correlated with $2s-1$.
Thus eigenvalues of a random state generated in this way
coincide with squared singular values of the product of $s$ independent 
random Ginibre matrices. Their statistical properties will be analyzed in the following section.
In general the Ginibre matrices need not to have the same dimension
so the model can be generalized. Assuming that a rectangular matrix $G_i$
has dimensions $N_i \times M_i$
one has to put $M_i=N_{i+1}$ for $i=1,\dots, s-1$,
so that the product  (\ref{prodrands}) is well defined.
Setting $N_1=N$ and defining the ratios $c_i=M_i/N$ for $i=1,\dots,s$ 
one obtains ensemble of random states parametrized by the
vector of coefficients, ${\bf c}:=\{c_1,\dots c_s\}$.

\begin{figure}[htbp]
\centering
\includegraphics[width=0.43\textwidth]{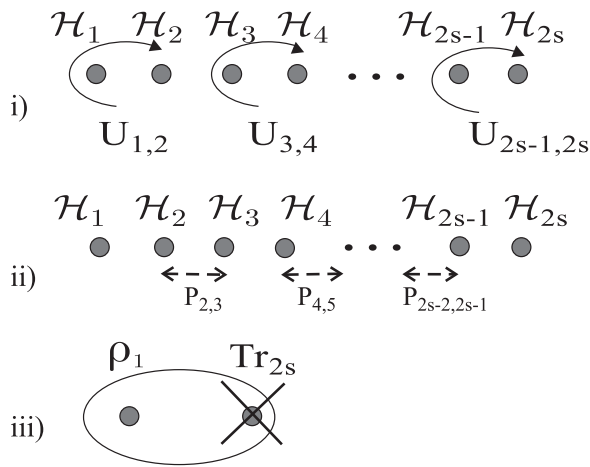}
\caption{To obtain a random mixed state with spectral density
       $\pi^{(s)}$ use a system consisting of $2s$ subsystems, $1,\dots, 2s$ of size $N$:
i) take the product of $s$ bi-partite random pure states generated by 
random unitary matrices $U_{1,2},U_{3,4},\dots U_{2s-1,2s}\in U(N^2)$, 
ii) measure subsystems  $2,\dots, 2s-1$ by a projection onto the
 product of $s-1$ maximally 
   entangled states,  
$P_{23}\otimes P_{4,5}\otimes \cdots \otimes P_{2s-2,2s-1}$
where $P_{i,j}=|\Psi_{i,j}^+\rangle \langle \Psi_{i,j}^+|$, 
iii) perform partial trace average over the subsystem $2s$.}
\label{fig6}
\end{figure}

\section{Product of Ginibre matrices and Fuss-Catalan distribution}

For any integer number $s$, there exists a probability measure $\pi^{(s)}$, 
called the Fuss-Catalan distribution of order $s$, 
whose moments are the generalized Fuss-Catalan  numbers  \cite{armstrong,mlotk}
given in terms of the binomial symbol,
\begin{equation}
\int_0^{b(s)} x^m \pi^{(s)}(x) dx = 
\frac{1}{sm+1}\binom{sm+m}{m} 
=:FC^{(s)}_m \ .
\label{FCmom}
\end{equation}

The measure $\pi^{(s)}$ has no atoms, it is supported on
 $[0,b(s)]$ where $b(s)=(s+1)^{s+1}/s^s$, its density is analytic on $(0,b(s))$,
 and bounded at $x=b(s)$, with asymptotic behavior $\sim 1/(\pi x^{s/(s+1)})$ at $x\to 0$.
This distribution arises in random matrix theory as one studies the product of 
$s$ independent random square Ginibre matrices, $W=\prod_{j=1}^s G_j$.
In this case squared singular values of $W$ (i.e. eigenvalues of
 $WW^{\dagger}$) 
have asymptotic distribution $\pi^{(s)}$. 
The same Fuss--Catalan distribution % $\pi^{(s)}$ 
describes asymptotically the statistics of singular values of $s$-th power
of a single random Ginibre matrix \cite{AGT10}.
In terms of free probability theory, 
it is the free multiplicative convolution product of $s$ copies of the 
Marchenko-Pastur distribution \cite{banica-etal,BG09}, 
which is written as $\pi^{(s)}=[\pi^{(1)}]^{\boxtimes s}$.

\begin{figure}[htbp]
\centering
\includegraphics[width=0.35\textwidth]{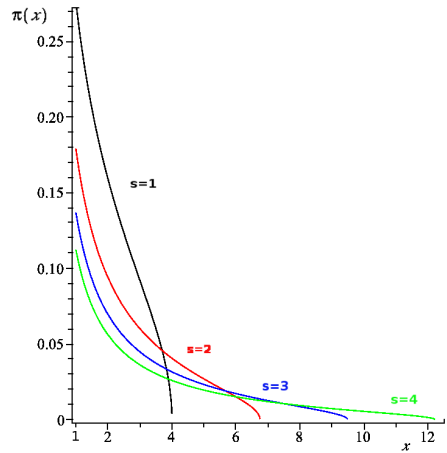}
\caption{Fuss--Catalan distributions $\pi^{(s)}(x)$
plotted for $s=1,2,3,4$ supported on the interval 
$[0,(s+1)^{s+1}/s^s]$.
To demonstrate the behavior at the right edge of the support
the figure is depicted for  $x \ge 1$.
}
\label{fig4g}
\end{figure}

An explicit expression of the spectral density for {$s=2$}, 
\begin{equation}
\!\!\pi^{(2)}(x) =  \frac{\sqrt[3]{2} \sqrt{3}}{12 \pi} \;
 \frac{\bigl[\sqrt[3]{2} \left(27 + 3\sqrt{81-12x} \right)^{\frac{2}{3}} -
   6\sqrt[3]{x}\bigr] } {x^{\frac{2}{3}}
     \left(27 + 3\sqrt{81-12x} \right)^{\frac{1}{3}}},
\label{pi2}
\end{equation}
where $ x \in [0,27/4]$,
was derived first in \cite{PS01} in context of 
construction of generalized coherent states from combinatorial sequences.
More recently it was applied in \cite{CNZ10} to describe random quantum states associated
with certain graphs.

The spectral distribution of a product of an arbitrary number 
of $s$ random Ginibre matrices was recently analyzed by Burda et al.  \cite{BJLNS10} 
also in the general case of rectangular matrices. The distribution was expressed
as a result of a polynomial equation and it was conjectured that the finite size effects
can be described by a simple multiplicative correction.
Another recent work of Liu et al.\cite{LSW10}
provides an integral representation of the distribution  $\pi^{(s)}$
derived in the case of $s$ square matrices of size $N$, which is assumed to be large.

Making use of the inverse Mellin transform and the Meijer $G$--function
one may find a more explicit form of this distribution.
It can be represented \cite{PZ10} as a superposition of $s$ hypergeometric functions 
of the type ${}_s F_{s-1}$,
\begin{widetext}

\begin{equation}
\label{eq:FCs}
\pi^{(s)}(x) = \sum_{n=1}^s  \Lambda_{n,s} \; x^{\frac{n}{s+1}-1}\; 
{}_sF_{s-1}\Bigl(\Bigl[\Bigl\{1-\frac{1+j}{s} +\frac{n}{s+1} \Bigr\}_{j=1}^s  \Bigr];
\Bigl[\Bigl\{1+\frac{n-j}{s+1} \Bigr\}_{j=1}^{n-1} ,
      \Bigl\{1+\frac{n-j}{s+1} \Bigr\}_{j=n+1}^{s} \Bigr];
         \frac{s^s}{(s+1)^{s+1}} x 
                   \Bigr)
\end{equation}
where the coefficients $\Lambda_{n,s}$ read for $n=1,2, \dots, s$
\begin{equation}
\label{eq:FCs2}
\Lambda_{n,s} := s^{-3/2} \sqrt{\frac{s+1}{2\pi}} 
 \Bigl(\frac{s^{s/(s+1)}}{s+1}\Bigr)^{n}  \;
 \frac{\Bigl[ \prod_{j=1}^{n-1} \Gamma\bigl(\frac{j-n}{s+1}\bigr) \Bigr]
       \Bigl[ \prod_{j=n+1}^{s} \Gamma\bigl(\frac{j-n}{s+1}\bigr) \Bigr] }
  { \prod_{j=1}^{s} \Gamma\bigl( \frac{j+1}{s} - \frac{n}{s+1}\bigr) } \ .
\end{equation}

Here ${}_pF_q\bigl( \bigl[ \{a_j\}_{j=1}^p\bigr] ;\bigl[ \{b_j\}_{j=1}^q \bigr];x\bigr)$
stands for the hypergeometric function \cite{MOT} of the type ${}_pF_q$
with $p$ 'upper' parameters $a_j$ and $q$ 'lower' parameters $b_j$
of the argument $x$. The symbol 
$\{a_i\}_{i=1}^r $ represents the list of $r$ elements, $a_1,\dots a_r$.
The above distribution is exact and it describes the density of 
squared singular values of $s$ square Ginibre matrices in the limit
of large matrix size $N$.

Observe that in the simplest case $s=1$ the above form reduces to
the Marchenko--Pastur distribution, 
\begin{equation}
\pi^{(1)}(x) = \frac{1}{\pi \sqrt{x}} 
\; {}_1F_0 \Bigl([-\frac{1}{2}];[ \ ];\frac{1}{4}x \Bigr) 
=\frac{\sqrt{1-x/4}}{\pi \sqrt{x}}\; ,
\label{pi1bis}
\end{equation}
while the case $s=2$ 
\begin{equation}
\pi^{(2)}(x) = \frac{\sqrt{3}}{2\pi x^{2/3}} 
\;{}_2F_1 \Bigl([-\frac{1}{6},\frac{1}{3}];[\frac{2}{3}];\frac{4x}{27} \Bigr) -
\frac{\sqrt{3}}{6\pi x^{1/3}} 
\;{}_2F_1 \Bigl([\frac{1}{6},\frac{2}{3}];[\frac{4}{3}];\frac{4x}{27} \Bigr)
\label{p21bis}
\end{equation}
is equivalent to the form (\ref{pi2})  obtained in \cite{PS01}.
\end{widetext}

The distributions (\ref{eq:FCs})
are thus directly applicable to describe the level density of
random mixed states obtained from a $2s$--partite pure states
by projection onto maximally entangled states and partial trace
as described in previous section.
This result becomes exact in the asymptotic limit,
it the dimension $N$ of a single subsystem tends to infinity.
However, basing on recent results of Burda et al.  \cite{BJLNS10} 
one can conjecture that the finite $N$ effects can be described by a 
multiplicative correction.

Note that the upper edge $b(s)=(s+1)^{s+1}/s^s$ 
of the FC distribution $\pi^{(s)}(x)$
for large matrices determines the size of the largest eigenvalue $\lambda_{max}$
of the reduced density matrix $\rho$ of size $N$. 
In the case of the structureless ensemble
of random pure states on $N \times N$ system, corresponding to $s=1$,
one has $b(1)=4$ so that $\lambda_{max}\approx 4/N$.
For states obtained by projection of a $2s$ partite system on maximally entangled states,
as described in the previous section, the largest component 
behaves as $\lambda_{max}\approx b(s)/N$.

The number  $\lambda_{max}$, equal to the  
largest component of the Schmidt vector of a random pure state 
on the bi--partite system, can be used to measure the degree of 
quantum entanglement. For instance, for a bi-partite system, 
the 'geometric measure' of entanglement, related to the 
distance to the closest separable state (in sense of the natural Fubini-Study distance)
reads \cite{WG03}, $E_g(|\phi\rangle) =-\ln \lambda_{\rm max}$.
 This quantity can also be considered as the Chebyshev entropy $S_{\infty}$ - 
the generalized Renyi entropy $S_q=\frac{1}{1-q} \ln {\rm Tr} \rho^q$
in the limit $q\to \infty$ \cite {BZ06}.

Thus the right edge $b(s)$ of the support of the spectral density for the reduced
state $\rho={\rm Tr}_B|\phi\rangle \langle \phi|$
determines the geometric measure of entanglement of the corresponding
random pure states. In the case of structureless random pure states,
related to the Marchenko--Pastur distribution one becomes  an asymptotic expression
$\langle E_g(|\phi\rangle) =-\ln (4/N)$. In the general case
of random state corresponding the the FC distribution $\pi^{(s)}(x)$
the typical value of the entropy reads
\begin{equation}
\langle S_{\infty}\rangle_s = \ln N -\ln b(s) = \ln N +s\ln s - (s+1) \ln (s+1).
\label{FCcheb}
\end{equation}
The larger value of $s$, the smaller the Chebyshev entropy $S_{\infty}$,
and the less entangled a typical random state
obtained by the projection of the $2s$ partite system.

The average von Neumann entropy, $S_1=-{\rm Tr}\rho\ln \rho$,
of mixed states of size $N$ generated according to the FC distribution reads
$\langle S(\rho) \rangle_s= \ln N - \sum_{j=2}^{s+1} \frac{1}{j}$ \cite{CNZ10}. 
The second moment of the FC distribution given in (\ref{FCmom}) implies
the asymptotic average purity $\langle {\rm Tr} \rho^2\rangle_{s}\approx (s+1)/N$
-- the larger number $s$, the more pure the typical mixed state generated
by a projection onto $s-1$ maximally entangled states.

\section{Concluding remarks}
\label{sec_concl}

In this work we analyzed structured ensembles of random pure states
on composite systems. They are defined with respect to a 
given decomposition of the entire system into its subsystems,
what induces a concrete tensor product structure in the Hilbert space.
The structured ensembles are thus invariant with respect to the
group of local unitary transformations.

Performing a partial trace over selected subsystems
one obtains an ensemble of random mixed states 
defined on the remaining subsystems. 
The particular ensemble depends thus on the number of systems
traced out and on the way the initial random pure states are prepared.

Quantum states obtained by the partial trace
of a superposition of $k$ maximally entangled pure states
of the bi-partite system involve the sum of $k$ 
random unitary matrices.
To generate states which involve a product of an arbitrary number
of $s$ matrices one needs to consider a system consisting of $2s$ subsystems,
 in which an orthogonal measurement is performed
in the product of $s-1$ maximally entangled bases.
Selected ensembles of random states, recipe to generate numerically
 the corresponding density matrices and some properties of the
distribution of the Schmidt coefficients are collected in Table 1.

We are going to conclude this work by writing down a more general class
of structured random states, which contains all particular cases 
discussed in the paper and listed in Table 1.
Consider the following ensemble of non-hermitian random matrices
parametrized by an arbitrary $k$--dimensional probability vector 
$p=\{p_1, \dots, p_k\}$
and a non-negative integer $s$, 
\begin{equation}
%W_{a,s} :=  \bigl[ {a\mathbbm 1}+(1-a)U \bigr] G_1 \cdots G_s.
W_{k,s} :=  \bigl[ p_1 U_1 +p_2 U_2 + \dots +p_k U_k \bigr] G_1 \cdots G_s .
\label{genera}
\end{equation}
Here $U_1, \dots U_k$ denote $k$ independent 
random unitary matrix distributed according to the Haar measure on $U(N)$,
while $G_1,\dots G_s$ are independent square random matrices of size $N$
from the complex Ginibre ensemble.
Random density matrix is obtained as a normalized Wishart--like matrix, 
\begin{equation}
\rho_{k,s} := \frac {W_{k,s}W_{k,s}^{\dagger}}{{\rm Tr} (W_{k,s}W_{k,s}^{\dagger})} .
\label{gensemble}
\end{equation}

Note that any particular ensemble from the above class 
can be physically  realized by taking a superposition of $k$
random pure states weighted by the vector $p$. 
Each pure state is defined on the system containing $2s$ subsystems. 
Performing a measurement in the product of $(s-1)$ maximally entangled bases %(\ref{projs})
one gets a random pure state of the desired structured ensemble. 
Eventually, averaging over the last subsystem 
one arrives at the mixed state (\ref{gensemble}).

\begin{widetext}

%%%%%%%%%%%%%%%%%%%%%%%%%%%%%%%%%%%%%%%%%%%%%%%%%%%%%%%%%%%%%%%%%%%%%%%%%%%%%%%%
% TABLE  I WITH  ensembles of random states
%%%%%%%%%%%%%%%%%%%%%%%%%%%%%%%%%%%%%%%%%%%%%%%%%%%%%%%%%%%%%%%%%%%%%%%%%%%%%%%%
\begin{table}[h]
\begin{tabular}{|c|c|c|c|c|c|c|c|}
\hline
\hline
% after \\: \hline or \cline{col1-col2} \cline{col3-col4}
$k$ & $s$ & matrix $W$ & distribution $P(x)$ & singularity at $x\to 0$
& support $[a,b]$ & $M_2$ & mean entropy \\
\hline
$1$ & $0$ & $U_1$   & $\delta(1) \; =\;  \pi^{(0)}$ & -- & $\{1\}$ & $1$ & $  0$ \\
$2$ & $0$ & $U_1+U_2$ & arcsine &  $x^{-1/2}$ & $[0,2]$ & $3/2$ & $  \ln 2-1\approx -0.307$ \\
$3$ & $0$ & $U_1+U_2+U_3$ & $3$ entangled states &  $x^{-1/2}$ & $[0,2\frac{2}{3}]$ 
&  $5/3$   & $ \approx -0.378$ \\
$4$ & $0$ & $U_1+U_2+U_3+U_4$ & $4$ entangled states &  $x^{-1/2}$ & $[0,3]$ 
&$7/8$   & $ \approx -0.411$ \\
$1$ & $1$ & $G\ \sim\  UG$ & Marchenko--Pastur $\pi^{(1)}$  &  $x^{-1/2}$ & $[0,4]$ & $2$ & $ -1/2=-0.5$ \\
$2$ & $1$ & $(U_1+U_2)G$ & Bures &  $x^{-2/3}$ & $[0,3\sqrt{3}]$ & $5/2$ & $ -\ln 2\approx -0.693$ \\
$1$ & $2$ & $G_1G_2$ & Fuss--Catalan $\pi^{(2)}$ &  $x^{-2/3}$ & $[0,6\frac{3}{4}]$ & $3$ & $ -5/6\approx -0.833$ \\
$1$ & $..$ & $... $ & $...$  & ... & ...& ... & ... \\
$1$ & $s$ &$G_1\cdots G_s$ & Fuss--Catalan $\pi^{(s)}$ &  $x^{-s/(s+1)}$ 
  & $[0,(s+1)^{s+1}/s^s]$ & $s+1$ & $ -\sum_{j=2}^{s+1} \frac{1}{j}$ 
\\ \hline \hline
\end{tabular}
\caption{Reduction of pure states from structured ensembles  
leads to random mixed states of the form  $\rho=WW^{\dagger}/{\rm Tr} WW^{\dagger}$.
Random matrix $W$ is constructed out of random unitary matrices $U_i$ distributed 
according to the Haar measure and/or (independent) random Ginibre matrices $G_j$
of a given size $N$. Asymptotic distribution $P(x)$ of the density of a
rescaled eigenvalue $x=N\lambda$ of $\rho$ for $N\to \infty$
is characterized by the singularity at $0$, its support $[a,b]$, the second moment $M_2$ 
determining the average purity $\langle {\rm Tr} \rho^2\rangle =M_2/N$
 and the mean entropy,
 $\int_a^b - x \ln x P(x) dx$, according to which the table is ordered.
} 
\label{tab1}
\end{table}
%%%%%%%%%%%%%%%%%%%%%%%%%%%%%%%%%%%%%%%%%%%%%%%%%%%%%%%%%%%%%%%%%%%%%%%%%%%%%%%%

\end{widetext}

Consider first the case of a uniform probability vector, $p_i=1/k$ for $i=1,\dots, k$.
For $k=1$ one obtains ensembles leading to the Fuss--Catalan distributions
$\pi^{(s)}$, which in the case $s=1$ reduces to the Marchenko--Pastur distribution.
Taking $k=2$ and $s=0$ one obtains the arcsine ensemble (\ref{cosrand}), 
while for larger $k$ one obtains the distributions (\ref{eq:density-k-rescaled}),
which converge to  $\pi^{(1)}$ in the limit $k \to \infty$.
Moreover, the case $k=2$ and $s=1$ corresponds to the Bures ensemble (\ref{burrand2}).
Thus the case $k=2$ and arbitrary $s$ can be called  {\sl higher order Bures ensemble}.

In a more general case, taking an arbitrary probability vector $p$
and varying the weights in a continuous manner one can study
transition between given structured ensembles.
For instance, by fixing the parameter $s$,  setting $k=2$ and varying the weight $p_2=1-p_1$
one defines a continuous interpolation between 
the higher order Bures ensemble and the Fuss--Catalan ensemble.
We have shown therefore that having in our disposal 
simple algorithms to generate random unitary 
and random Ginibre matrices we can construct
a wide class of ensembles of random quantum states.
Furthermore, we provide constructive physical recipe to generate 
such states by means of generic two-particle interaction,
superposition of states, selective measurements in maximally entangled basis
and performing averages over certain subsystems. 

As discussed in Appendix B it is also possible to introduce
 analogous ensembles of real random density matrices.
Physically this corresponds to imposing restrictions on 
the class of the interactions used to generate random pure states. 
In contrast with the complex case, 
the ensemble based on square real Ginibre matrices
does not lead to the Bures measure in the space of real states.
To achieve such a measure one needs to generalize the ensemble even further
to allow also rectangular Ginibre matrices \cite{OSZ10}.
In physical terms this implies that the dimension 
of the principal system and the auxiliary system have to be different in this case.

The notion of random quantum states is closely related with
the concept of random quantum maps.
Due to the Jamio{\l}kowski isomorphism any quantum operation $\Phi$
acting on density matrices of size $N$
can be represented by a state on the extended Hilbert space \cite{BZ06},
\begin{equation}
\sigma = \bigl( \Phi \otimes {\mathbbm 1} \bigr) |\psi^+\rangle \langle \phi^+|.
\label{jamiol}
\end{equation}
Here  $|\psi^+\rangle = \frac{1}{\sqrt{N}} \sum_{j=1}^N |j,j\rangle$
denotes the maximally entangled state from the bipartite Hilbert space
${\cal H}={\cal H}_N \otimes {\cal H}_N$.
Any state $\sigma$ on the composed Hilbert space
${\cal H}={\cal H}_A \otimes {\cal H}_B$
 defines a completely positive, trace preserving map
provided the following partial trace condition is satisfied,
${\rm Tr}_A \sigma({\Phi})  =  {\mathbbm {1}}/N$.

It is possible to impose this partial trace condition on an arbitrary 
 state $\omega$ acting on $\cal H$.
To this end one finds the reduced state $Y:={\rm Tr_A} \omega$
 which is positive and allows one to take its its square root $\sqrt{Y}$,
and writes the normalized cognate state \cite{BCSZ09},
\begin{equation}
\sigma=
\frac{1}{N} \Bigl(  {\mathds 1} \otimes \frac{1}{\sqrt Y} \Bigr)
 \omega
\Bigl(  {\mathds 1}  \otimes \frac{1}{\sqrt Y}\Bigr)
 \ ;
\label{Dren}
\end{equation}
The required property, ${\rm Tr}_A \sigma  =  {\mathbbm {1}}/N$,
is satisfied by construction, so the state $\sigma$ 
represents a quantum operation. 
As the matrix elements of the corresponding  superoperator $\Phi$
can readily be obtained by reshuffling \cite{BZ06} the entries of the density matrix $\sigma$,
any random state $\omega$ determines  by (\ref{Dren}) and (\ref{jamiol}) a quantum operation.
Therefore  any ensemble of random states introduced in this paper, 
applied for bi-partite, $N \times N$ systems determines 
the corresponding  ensemble of random operations.
 For instance, the induced measure with $K=N^2$ 
corresponds to the flat measure in the space of quantum operations \cite{BCSZ09,BSCSZ10},
but other ensembles of random states can be also applied to generate 
random  quantum operations \cite{BZ06}.

\medskip 

The present study on ensembles of  random states should be
concluded with a remark that apart of the methods developed in this paper
several other approaches are advocated in the literature.
In very recent papers \cite{PFPPS10,NMV10}
the authors follow a statistical approach introducing 
a partition function which leads to a generalization of
the Hilbert-Schmidt measure. Varying the parameter of the model,
which corresponds to the inverse temperature, they demonstrate 
a phase transition during an interpolation between Marchenko-Pastur
and semicircle distribution of spectral density. 
In  another recent approach Garnerone et al. study statistical properties
of random matrix product states \cite{GOHZ20}, which are obtained 
out of products of truncated random unitary matrices. 
Although these models of random states do differ from the one 
presented in this work, a possible links and relations between 
results obtained in these approaches is currently
under investigation.

\medskip

Acknowledgments.
It is a pleasure to thank M.~Bo{\.z}ejko, Z. Burda,
 K.J.~Dykema, P.~Garbaczewski, V.A.~Osipov,  W.~M{\l}otkowski, 
H.-J.~Sommers, F. Verstraete, W. Wasilewski and  M. {\.Z}ukowski 
for several fruitful discussions and to T. Pattard for a helpful correspondence.
We are thankful to F. Benatti,  J.A. Miszczak and to an anonymous referee, 
whose remarks helped us to improve the paper considerably.
% on ensembles of random density matrices.
Financial support by the Transregio-12 project 
der Deutschen Forschungsgemeinschaft and
the grant number  N N202 090239 of
Polish Ministry of Science and Higher Education
is gratefully acknowledged.
KAP acknowledges support from Agence Nationale de la Recherche (Paris, France)
under program No. ANR-08-BLAN-0243-02.
BC and IN were supported by BC's 
NSERC discovery grants, BC acknowledges  partial
support from the ANRs GRANMA and GALOISINT,
IN was supported by BC's Ontario's Early Research Award. 
K{\.Z} acknowledges the support of the PI workshop 
``Random Matrix Techniques in Quantum Information Theory'', 
during which BC, IN and K{\.Z} worked together on this paper.

\appendix

\section{Sum of $k$ random unitaries and the distribution $\nu_k$}
\label{sec_karand}

We compute, in the limit of large matrix size $N \to \iy$, the asymptotic eigenvalue distribution of the random matrix 
\begin{equation}
	V_k = \frac{1}{k}(U_1 + U_2 + \cdots + U_k)(U_1^* + U_2^* + \cdots + U_k^*),
\end{equation}
where $U_1, \ldots, U_k$ are $N \times N$ random independent Haar unitary matrices.
 Obviously, this is equivalent to computing the singular value distribution of $k^{-1/2}\sum_{i=1}^k U_i$. 

For now, we forget about the normalization pre-factor and we put $W_k = kV_k$. 
It is a well known result in free probability theory that independent large unitary matrices are free from each other 
(and, very importantly but not of interest here, they are also free from deterministic diagonal matrices):

\begin{theorem}
[\cite{voiculescu-dykema-nica} and \cite{nica-speicher}]

 Let $U_{1, N}, \ldots, U_{k,N} \in \U(N)$ be $k$ independent Haar unitary random matrices.
 Then, as $N \to \iy$,
\begin{equation}
	U_{1, N}, \ldots, U_{k,N} \xrightarrow{*-\text{distr}} u_1, \ldots, u_k,
\end{equation}
where $u_1, \ldots, u_k$ are free Haar unitary elements in a non-commutative $W^*$-probability space $(\M, \tau)$.
\end{theorem}

Hence, computing the limit distribution of $W_k$ amounts to understanding the distribution of a sum of 
$k$ free Haar unitary elements in a von Neumann algebra $w_k = (u_1 + \cdots + u_k)(u_1^* + \cdots + u_k^*)$. 
This problem has been related to random walks on $k$-regular trees by Kesten \cite{kesten}. 
Indeed, the number of alternating words of length $2p$ in the letters $u_i, u_i^*$ which reduce 
to the unit is bijectively the same as the number of walks of length $2p$ on the $k$-regular tree beginning and ending at some fixed vertex. 
Using standard formulas for the number of such walks, we can deduce moment information for $k=2$
\begin{equation}
\lim_{N \to \iy} \E\left[ \frac{1}{N} \trace W_2^p \right] = \tau(w_2^p) = \binom{2p}{p}^2,
\end{equation}
and a moment generating function in the general case
\begin{equation}
	F_k(z) = \sum_{p=0}^\iy \tau(w_k^p)z^p = \frac{2(k-1)}{k-2 + k \sqrt{1-4(k-1)z}}.
\end{equation}

From the last formula, using Cauchy transform techniques, one can easily deduce the probability density function of the distribution of $w$:
\begin{equation}\label{eq:density-k}
	d\mu_k(x) = \frac{k}{2\pi} \frac{\sqrt{4(k-1)x - x^2}}{k^2x - x^2} \ind_{[0, 4(k-1)]}(x) dx.
\end{equation}

This result has been obtained by Haagerup and Larsen in \cite{haagerup-larsen}, Example 5.3. The authors were interested in the Brown measure of the non-normal element $\tilde w = u_1 + \cdots + u_k$. They found that the distribution of $|\tilde w|$ is given by
\begin{equation}
d\tilde \mu_k(x) = \frac{k}{\pi} \frac{\sqrt{4(k-1) - x^2}}{k^2 - x^2} \ind_{[0, 2\sqrt{k-1}]}(x) dx.	
\end{equation}
One can easily recover equation \eqref{eq:density-k} by using $w = \tilde w \tilde w^*$ and by noticing that $\mu_k = \tilde \mu_k \circ \mathrm{sq}$, where $\mathrm{sq}$ is the square function $\mathrm{sq}(x) = x^2$.

If $\nu_k$ is the distribution of the rescaled element $v=w/k$, then one 
arrives at the desired distribution function,
\begin{equation}
\label{eq:density-k-rescaled2}
	d\nu_k(x) = \frac{1}{2\pi} \frac{\sqrt{4k(k-1)x - k^2x^2}}{kx - x^2} \ind_{[0, 4\frac{k-1}{k}]}(x) dx.
\end{equation}

\medskip

In \cite{haagerup-larsen}, it is shown that the Brown measure of $\tilde w$ is 
a rotationally invariant measure, supported on the centered disk of radius $\sqrt k$ with radial density
\begin{equation}
	f_{\tilde w}(r) = \frac{k^2(k-1)}{\pi(k^2 - r^2)^2}, \qquad  0 < r < \sqrt k.
\end{equation}
With the proper $k^{-1/2}$ rescaling, it is easy to see that the above
 Brown measure converges to the uniform measure on the unit disk. 
Hence, we recover the Ginibre behavior in the limit $k \to \iy$ 
(one has to take first the limit $N \to \iy$).
Let us add that a more general study on statistical properties of a
sum of random unitary  matrices was recently presented by Jarosz \cite{Ja10}.

\section{Real random states, real Ginibre and random orthogonal matrices}
\label{sec_real}

Although most often one considers complex density matrices,
it is also interesting to study quantum states described by real 
density matrices. The dimensionality of the set of real states 
on ${\cal H}_N$ is $N(N+1)/2-1$,
so its geometry is easier to study than that of the $N^2-1$ dimensional set of
complex states \cite{BZ06}. For instance, the set of real states of a qubit
forms a two dimensional disk, which can be considered as a cross-section
of the three dimensional Bloch ball of complex states.
Euclidean volume of the set of real density matrices of size $N$
was derived in \cite{ZS03}, while the corresponding measure 
can be derived from the real Ginibre ensemble.

In this appendix we define ensembles of real states 
based on random orthogonal matrices and real Ginibre ensemble
and show that the level density does not differ from the complex case.
To this end we formulate two lemmas.

\begin{lemma}
 Consider $k$ independent orthogonal matrices $O_1,\dots O_k$,
distributed according to the Haar measure on $O(N)$ and define a
normalized real density matrix
\begin{equation}
\rho_{\rm ort} \ = \ 
 \frac{ (O_1 + \dots +O_k)(O_1^T + \dots +O_k^T)}
{{\rm Tr} (O_1 + \dots +O_k)(O_1^T + \dots +O_k^T)} .
\label{cosortk}
\end{equation}
Then for large $N$ its spectral density is described 
by the distribution $\nu_k$ given in 
(\ref{eq:density-k-rescaled}).
\end{lemma}

This lemma follows directly from the fact that the moments 
of orthogonal and unitary random matrices have
 the same behavior for large matrix size $N$, 
since random orthogonal matrices are asymptotically free \cite{CS06}.
 
\medskip
\begin{lemma}
 Consider $s$ independent random matrices $R_1,\dots R_s$
taken from the real Ginibre ensemble of square matrices of size $N$.
Define  a normalized real density matrix
\begin{equation}
\rho_R \ = \ 
 \frac{ R_1 R_2 \cdots R_s (R_1 R_2 \cdots R_s)^T}
{{\rm Tr}[ R_1 R_2 \cdots R_s (R_1 R_2 \cdots R_s)^T ]}
\label{realgin}
\end{equation}
Then for large $N$ its spectral density is described 
by the Fuss-Catalan distribution $\pi^{(s)}$ given in 
(\ref{eq:FCs}).
\end{lemma}.

To prove this one needs  to show that in the case of large 
matrix size $N$ the moments of this distribution 
are indeed given by the Fuss--Catalan numbers  (\ref{FCmom}),
exactly as for the product of complex  Ginibre matrices.
This follows from the  interpretation of the Wick formula 
for random matrices in terms of maps gluing
and from the fact that leading terms have to be non-crossing, 
therefore orientable as for complex Gaussian matrices - see e.g. \cite{Zv97,Ok00}.

 \medskip

It is natural to combine both definitions
and defining a more general ensemble of real
density matrices, each obtained out of $k$ random
orthogonal matrices and $s$ square real Ginibre matrices 
in a direct analogy to eq. (\ref{genera}).

Let us close this section with a remark that
the differences between the real and complex case 
can be significant in some cases.
For instance, to get a real state distributed according to the
Bures measure one needs to use a symmetric random unitary matrix 
and a rectangular, $N\times (N+1)$, real Ginibre matrix, 
while the complex Bures state is obtained from a random unitary
and a square matrix of the complex Ginibre ensemble \cite{OSZ10}.

\end{document}